\documentstyle[iopconf,epsfig,epic]{article}
\newcommand{\slB}{ B\!\!\!\!\slash}
\newcommand{\slC}{\raise.15ex\hbox{$/$}\kern-.57em\hbox{$C$}}
\newcommand{\slCP}{C\raise.15ex\hbox{$/$}\kern-.57em\hbox{$P$}}
\newcommand{\be}{\begin{equation}}
\newcommand{\ee}{\end{equation}}
\newcommand{\lsi}{\raise0.3ex\hbox{$<$\kern-0.75em\raise-1.1ex\hbox{$\sim$}}}
\newcommand{\gsi}{\raise0.3ex\hbox{$>$\kern-0.75em\raise-1.1ex\hbox{$\sim$}}}

\begin{document}
\title{The Electroweak Phase Transition -- Standard and ``Beyond''
\footnote{Talk presented at the workshop ``Beyond the Desert'',
Accelerator and Nonaccelerator Approaches, Castle Ringberg, Tegernsee,
June 8-14, 1997}}
\author{Michael G. Schmidt}
\affil{Institut f\"ur Theoretische Physik\\ Universit\"at Heidelberg\\
Philosophenweg 16, D-69120 Heidelberg, Germany}
\beginabstract

We first review why a strongly first-order phase transition
needed for baryogenesis is excluded in the electroweak
standard model. We also
comment on some intriguing effects in the strongly
interacting hot phase. In the MSSM with a light stop
a strongly first-order phase transition can be
achieved. It possibly proceeds in two stages.
\endabstract

\section{Introduction}
Heating up the electroweak matter of the standard model (SM) a phase
transition (PT) is expected \cite{1} at a temperature $T_c\sim 10^{15}
K\ (\sim 100$ GeV) since the positive plasma mass$^2\sim(g_wT)^2$
of the Higgs field switches the sign of the Higgs
field mass$^2$. In
this simple picture the Higgs mechanism is suspended at
high temperatures, the Higgs-field VeV becomes zero; naively
in the ``hot phase'' the transversal $W$-bosons would be massless.

Such a PT should have occurred in the early universe at about
$10^{-12}$ sec after the big bang. Most of the interest in it in the
last years came from the observation that the electroweak
interactions violate the baryon number $B$: At $T=0$ the
instanton tunneling between topologically different vacua
related to $B$-violation \cite{2} is an immeasurably small effect
$\sim e^{-8\pi^2/g^2_w}$ (unless perhaps strongly
enhanced in multi-$W$ production \cite{3}). For $T$ below $T_c$
a $B$-violating thermodynamical transition between states with
neighbouring topological quantum numbers $N_{cs}$ via an
unstable 3-dimensional sphaleron configuration \cite{4,5} is
Boltzmann-suppressed and has a rate/volume
\be
\label{eq1}
\Gamma^{\rm Higgs\ Phase}_{ B\!\!\!\!\slash}\sim(\alpha_wT)^4e^{-S_3/T}.\ee
$S_3$ is the sphaleron action, which can be rescaled as
$v(T)/g_w \cdot 2\pi A$ where $v(T)$ is the $T$-dependent
Higgs VeV and $A$ a number $(\sim 3)$ only slowly varying with the
electroweak parameters.

For the large $T$ of
the hot phase it is argued
that there is no Boltzmann suppression and that
\be\label{eq2}
\Gamma^{\rm Hot\ Phase}_{\slB}\simeq K \alpha_w^n T^4.\ee
(The prefactor $K$ and the power $n$ are under discussion \cite{6}).

Baryon minus lepton number $B-L$ is conserved in electroweak
interactions.
If it is zero in the very early universe, it stays zero
cooling down to the PT.  During the equilibrium period
before the PT
$B$ as well as $L$ is washed out in this case. The observed baryon asymmetry
$\Delta B/n_\gamma\sim 10^{-10}$ should then be produced during
the PT. Indeed the three necessary criteria of Sakharov \cite{7}
for producing a $B$-asymmetry may be fulfilled in the SM \cite{8}:
(i)baryon-number violation, (ii) $C,\ CP$ violation, and
(iii) nonequilibrium. Concentrating
on the last point in this talk we see that a first-order PT is needed.

The most attractive way to produce a $B$-asymmetry is the
``charge transport mechanism'' \cite{9}. The walls of the
expanding bubbles of condensing electroweak
matter contain a $C,\ CP$ violating
phase factor and like a diaphragm produce axial
charge which is transported
into the hot phase in front of the bubble. There it is converted into
a baryonic charge
by the ``hot sphaleron''-transition (2)
before the Higgs phase bubble
takes over. Thus first the B-asymmetry should
be produced in the PT, then it should quickly ``freeze out''
in the Higgs phase. Therefore the baryon number violating rate
in the Higgs phase
should be small compared to the inverse Hubble
time $H\sim T^2_c/m_{Planck}\sim e^{-40}
T_c$ (for $T_c\sim 100 $ GeV) from where
it follows \cite{8} that $v(T_c)/T_c \geq1.3$ in the Boltzmann factor in (\ref{eq1}).
Thus one needs a strongly first-order PT.

\section{The electroweak phase transition in the SM}
To learn about the electroweak PT one conveniently inspects an
equilibrium quantity, the effective Higgs potential. The weak coupling
$g_w$ is small but at high temperatures $T$ the relevant expansion
parameter $\bar g^2_3$ is the 3-dimensional gauge coupling$^2$
$g_3^2\sim g_w^2T$ divided by some infrared (IR) scale $m_{IR}, \bar
g_3^2=g^2_3/m_{IR}$, and is not small for small $m_{IR}$. The most
elegant way of separating the infrared problem is the reduction from a
4- to a time-independent 3-dimension effective theory containing the
light degrees of freedom. Restricting the form of the
effective action (derivative expansion) the ``matching'' \cite{10, 11}
of 4- and 3-dimensional amplitudes allows to fix its parameters. This
procedure is purely perturbative and has been carried through to
2-loop order.  This loop order is necessary if one wants a few percent
accuracy $(0(g_w^3))$ \cite{12}. In a first step all $n\not=0$
Matsubara modes with $m_n=2\pi nT$ thus including all fermion modes
($n$ half integer) are integrated out in the above sense.In a second
step also the longitudinal gauge bosons which have obtained a Debye
mass $\sim g_wT$  in the first step are integrated out. One
ends up with the effective Langrangian 
\be\label{eq3}
L_3^{eff}=\frac{1}{4}F_{ik}^{a2}+(D_i^wH)^+(D_i^wH) +m^2_3
H^+H+\lambda_{H_3}(H^+H)^2.
\ee 
The U(1) part and Weinberg mixing can
be neglected in this discussion without loosing an essential point.
The SU(2)-Yang-Mills Lagrangian and the covariant derivative $D^w_i$
contain the gauge coupling $g^2_3=g_w^2T(1+...)$. The
Higgs parameters $m^2_3(T)$ and $\lambda_{H_3}(T)$ depend on the
temperature.
They can be made dimensionless in the ratios 
\be\label{eq4}
y=\frac{m_3^2(T)}{(g^2_3)^2},\quad x=\frac{\lambda^2_{H_3}(T)}
{g^2_3}.
\ee 
$y$ is related to $T-T_c$ and
$x(\sim \lambda_T/g_w^2$ in terms of 4-dimensional quantities)
determines the nature of the phase transition. $L_3^{eff}$
characterizes a whole class of theories. The specific properties of
the 4-dimensional theory only enter via the computation of $y, x$.

$L_3^{eff}$ as it stands as a tree level theory would give
a 2nd-order PT. But of course it has to be studied in higher
perturbative order and more than that it has
to be treated as a potentially  strongly
interacting QFT like QCD because of its IR behaviour.
Thus the most secure way is to discretize it on a lattice and to
discuss the results of lattice calculations \cite{12,13}. Still
perturbation theory can provide some interesting
insights. The simple one-loop $W$-boson exchange graph 
in a constant Higgs field background $\phi$
(3-dimensional, with $\phi_4=T^{1/2}\phi$) shown below
contributes the well-known term
\be\label{eq5}
V_3^{\phi^3}=-\frac{1}{24\pi}(g_w^2T\phi^+\phi)^{3/2}
=-E(T\phi^+\phi)^{3/2}\ee
\begin{center}
\epsfig{file=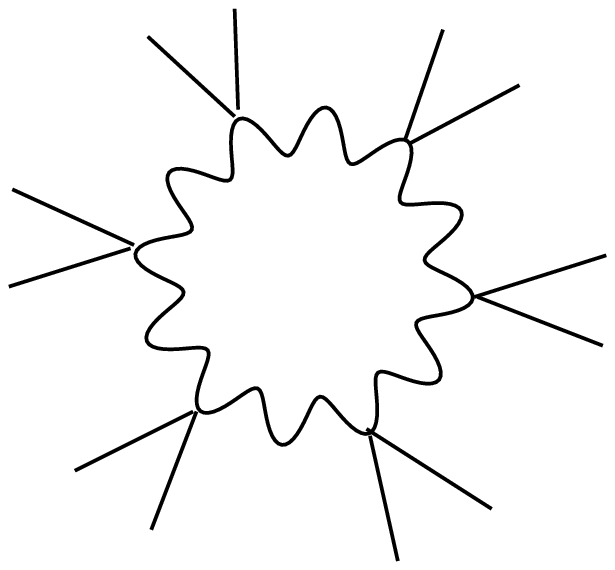,width=3cm}
\end{center}
to $V_3^{eff}$
leading already to a first-order PT. At the Higgs phase minimum
$v(T)$ perturbation theory in 2-loop order (figs. 1,2) indeed compares
very well with lattice results if $\bar g_3^2=g^2_3/v(T)
\sim cx$ is sufficiently small 
\cite{12,13}. Even the critical temperature
$T_c$ which one obtains by comparing the Higgs $\phi
=v(T)$ and the $\phi=0$ IR sensitive
minimum, can be determined quite
reliably this way for $x\lsi 0.08$. Most
sensitive is the surface tension of the critical bubble where we
observe \cite{12}, \cite{15} (fig. 3) strong deviations of the perturbative
calculation from lattice results for $x\gsi
0.05$.

\begin{figure}
\begin{minipage}{\textwidth}
\centering
\epsfig{file=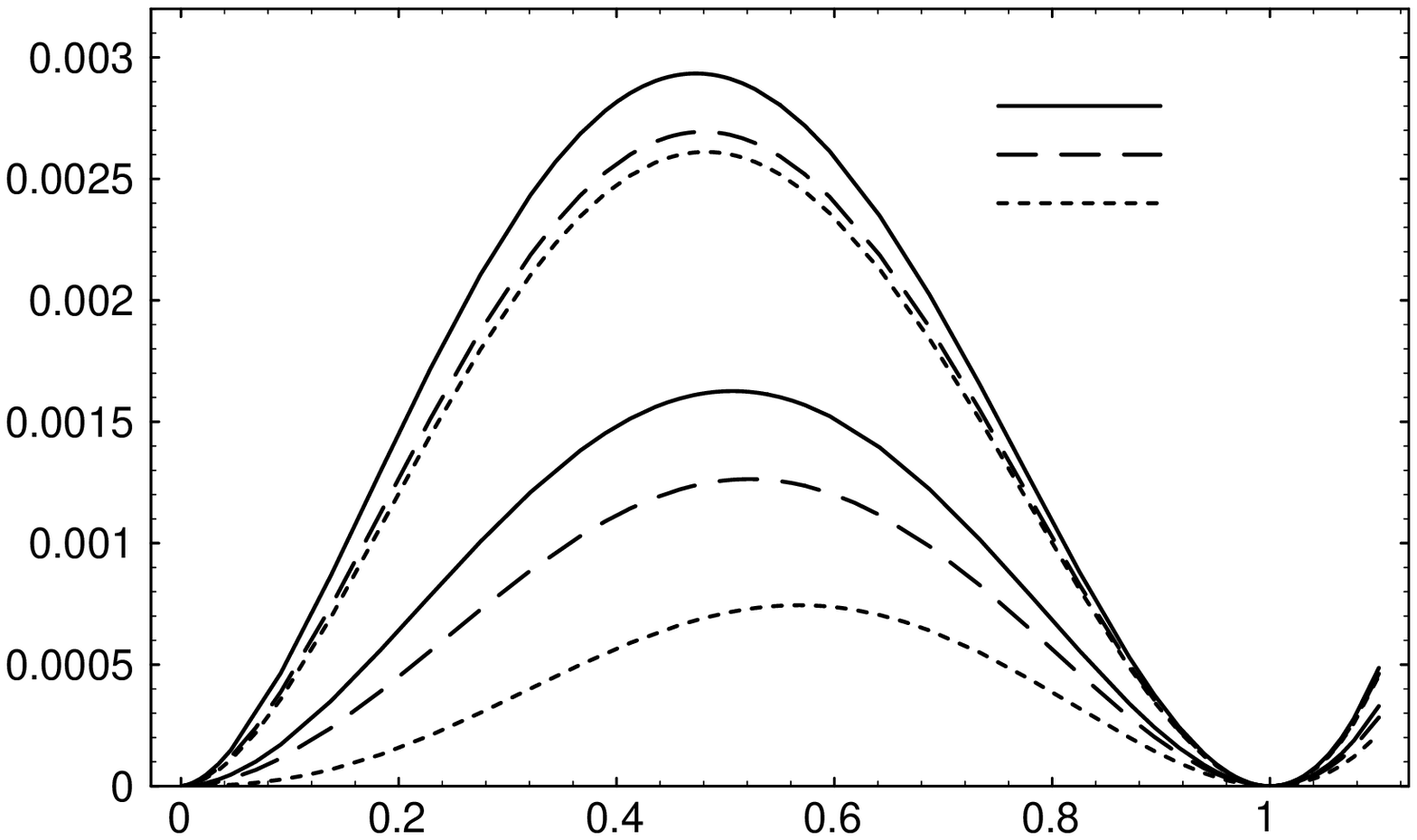,width=10cm}
\put(-68,8) {$\varphi/v$}
\put(-300,150){$\frac{V_{eff}}{g_w^2 v^4}$}
\put(-50,152) {$\xi=0$}
\put(-50,142) {$\xi=1$}
\put(-50,132) {$\xi=2$}
\caption{The 1- (lower) and 2-loop (upper curves) effective
potential at $T_c$ for $x=0.12$ in units of $v(T)$ in different
$\xi$-covariant background gauges (from ref. [14]). The
phase transition becomes stronger and the gauge dependence
diminishes in the 2-loop results.}
%
\centering
\epsfig{file=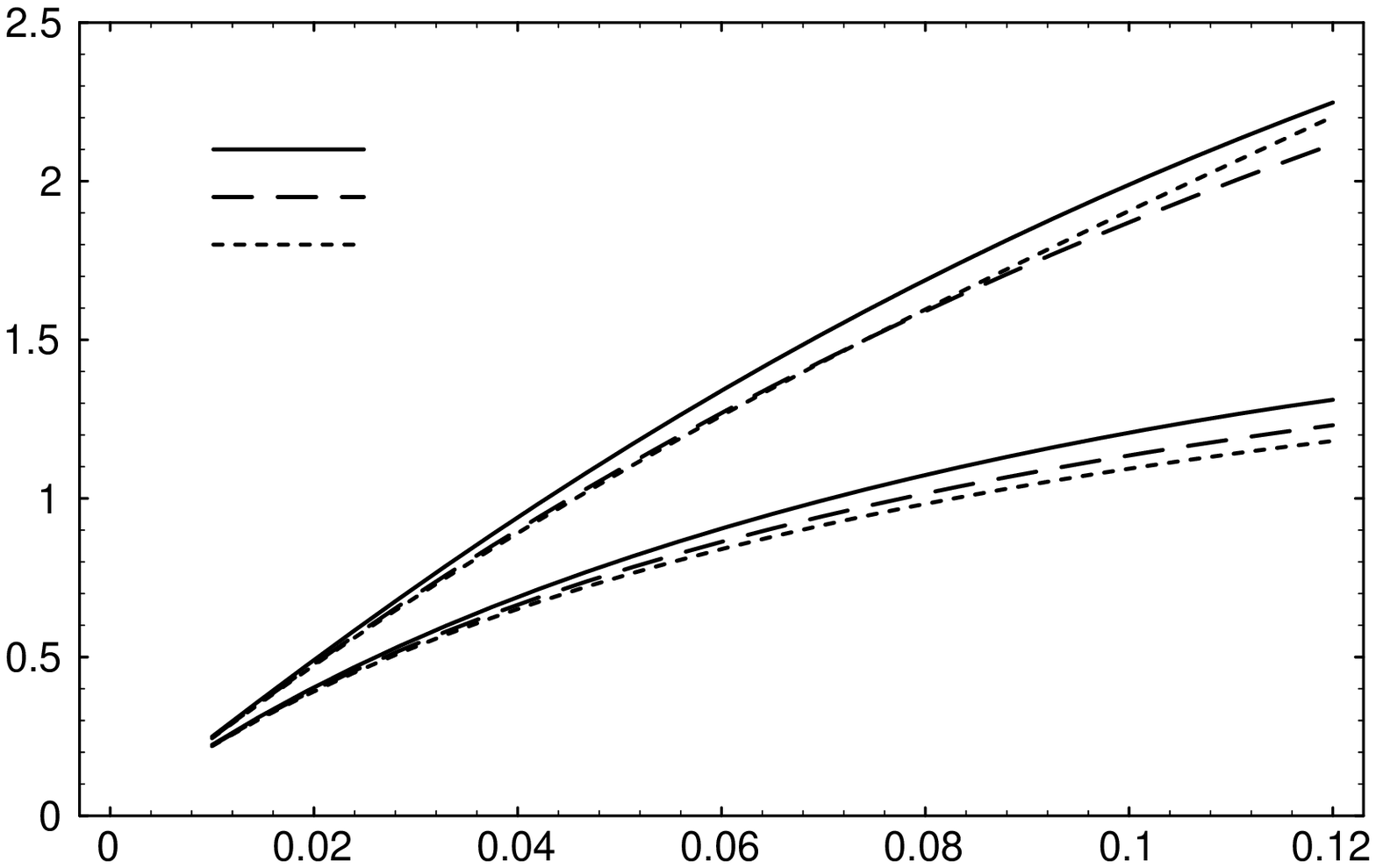,width=10cm}
\put(-45,0) {$x$}
\put(-300,150){$\frac{g_3^2}{g_w v}$}
\put(-203,147) {$\xi=0$}
\put(-203,137) {$\xi=1$}
\put(-203,127) {$\xi=2$}
\caption{$g^2_3(T_c)/(g_wv(T_c)) \sim g_wT_c/v(T_c)$ as a function of $x$
(1-loop: upper, 2-loop: lower curves)
(ref. [14]); for $x>0.04$ one
has $v(T_c)/T_c<1$.}
\end{minipage}
\end{figure}
\begin{figure}[ht]
\begin{minipage}{\textwidth}
\centering
\epsfig{file=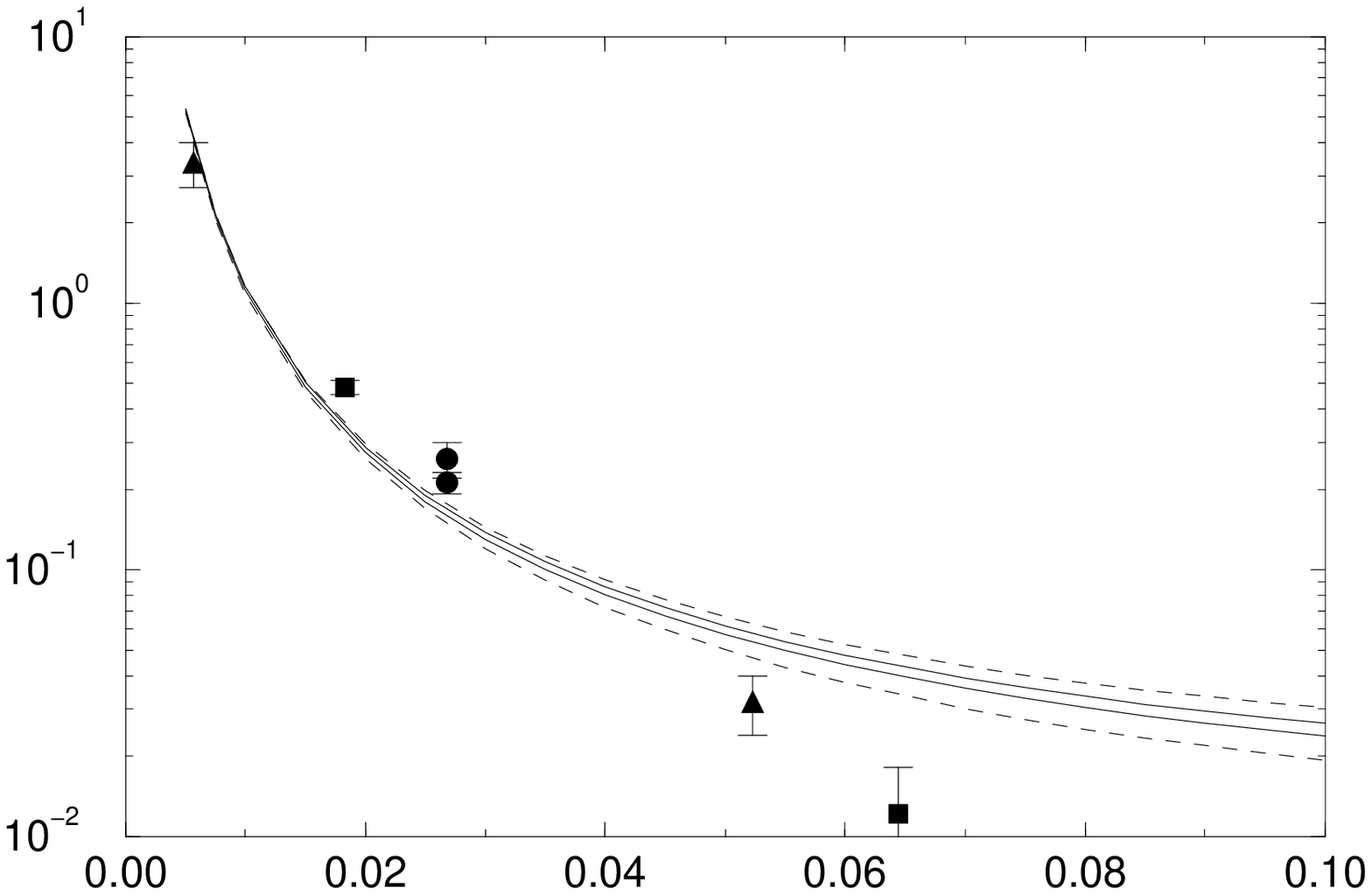,width=10cm}
\put(-45,15) {$x$}
\put(-280,150){$\frac{g_w^2 \sigma}{\left(g_3^2 \right)^3}$}
\caption{from ref. [15]) The perturbatively calculated
interface tension $\sigma$ (including $Z$-factor effect
and gauge variations)
vs. $x$ compared to lattice data from ref. [12] (squares),
ref. [16] (triangles) and ref. [17] (circles).}
\centering
\epsfig{file=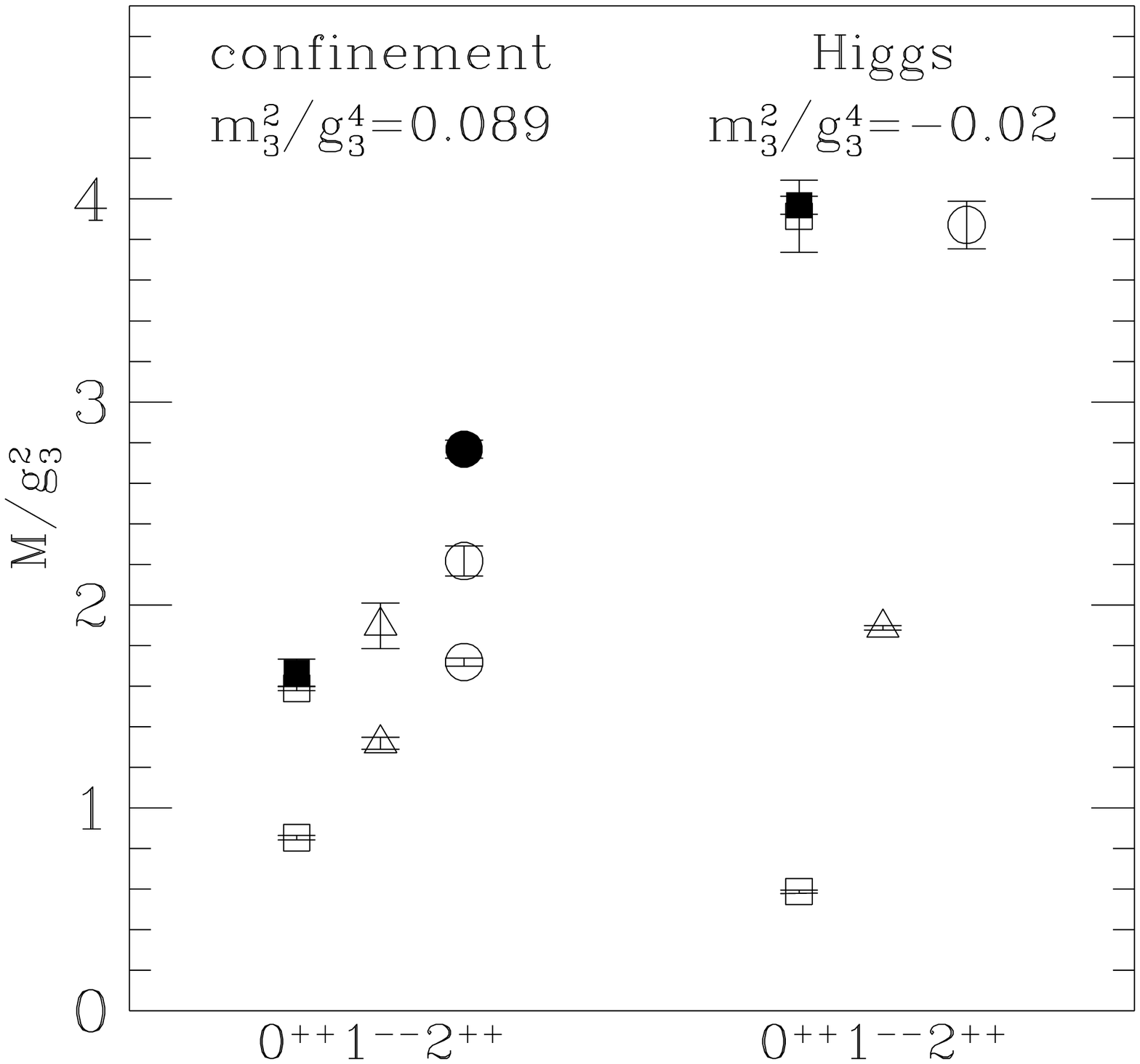,width=7cm}
\caption{(ref. [20]) Lattice results for the correlation masses for $0^{++}$ 
($J^{PC}$), $1^{--}$ and $2^{++}$ operators (at $x=0.0239$).
Dark points indicate purely "gluonic" operators.
Whereas in the Higgs phase only $H$, $W$ and multiple
(2$W$,...) are seen, in the hot phase one observes confinement 
and a completely different massive spectrum.}
\end{minipage}
\end{figure}
An effective potential $V^{eff}(\phi)$ and also the $Z$-factor
of the kinetic term is needed if one wants to calculate critical
bubbles \cite{15a}
and sphaleron field configurations \cite{5}. Thus some perturbative
expression enlarged by some nonperturbative
piece is desirable.

Most excitingly, lattice calculations \cite{15} show a
``crossover'' behaviour (qualitatively predicted in ref. \cite{16a},
\cite{16b})
in this phase diagram for
$x\gsi 0.11$ (corresponding to
$m_H\gsi m_W)$:
The first order PT fades away at such values of $x$. Of course a strongly
first-order PT with $v(T_c)/T_c>
1$ avoiding the sphaleron erasure of $B$-asymmetry requires
$x\lsi 0.03-0.04$ much below the crossover.
Indeed, a careful perturbative analysis of $x$ in terms of the physical
Higgs mass $m_H$ and the top mass gives \cite{10}
\be\label{eq5a}
x\sim\frac{1}{8}\frac{m_H^2}{m_W^2}+c\frac{m^4_{top}}{m^4_W}\ee
and shows that the above 
limit excludes  SM baryogenesis for any $m_H$!

Still the analysis of the SM at high $T$ based on the Lagrangian (\ref{eq3})
and its lattice regularization leads to results very interesting by
themselves: In the hot phase there is a confining linear potential
-- about the same as in pure YM theory \cite{13}. On the lattice
one can measure a rich spectrum (fig. 4) of $W$-balls and of ``Higgs hadrons''
which are QCD-type bound states of Higgses rather than quarks.
One should keep in mind that the masses of these bound states are 
3-dimensional correlation masses. These
masses can also be calculated \cite{17} in a
relativistic bound-state model with a confining
potential and compare very
well with lattice results. One can call spin 0
and spin 1 states Higgses and massive $W$-bosons respectively,
but one should emphasize that the hot phase is not another
Higgs phase although there are no massless vector bosons as in
the naive picture mentioned in the introduction. It would be
interesting to have a complete model of these bound states
in the whole phase diagram ($y$ versus $x$) and to compare it with
lattice results. It is also desirable to have a concrete
model \cite{18a}
of how the perturbative effective potential is modified
by a nontrivial vacuum structure including gauge-field condensates
at small values of $\phi$. This is particularly
important if the perturbative potential and its Higgs
minimum are small as they are in the case of $x$ values in the
crossover region and beyond.\clearpage

We conclude that the high $T$ electroweak standard theory
does not provide a first-order phase transition which is
strong enough for baryogenesis and that it even vanishes
for $m_H\gsi m_W$. It is also
very questionable if standard CP violation is large
enough. Still a lot of know-how also concerning $B$-asymmetry
production has accumulated. Thus if one does not want
to go back to $(B-L)$ violating asymmetry production in GU theories,
it might be attractive to consider  variants of the SM.

\section{Variants of the SM, the MSSM with a light stop}
It is widely accepted that the SM is an effective theory
to be embedded in a deeper theory including the Planck
scale. It is not clear, however, if at all and how it should
be varied at the electroweak
scale considering the great experimental
success of the SM. Taking a pragmatic view baryogenesis at
the electroweak scale requires a strongly first-order PT with $
v(T_c)/T_c>1$.
In 1-loop order one has $v(T)/T\sim E/\lambda_T$, where $E$
is defined in eq. (\ref{eq5}). This can be increased by\\
(i) increasing the E-prefactor of the ``$\phi^3$-term'' having more
light bosons in the loop. (below (\ref{eq5}))\\
(ii) decreasing the coupling $\lambda_T=\lambda_3/T$.\\
There
are further points to be mentioned:\\
(iii) The 2-loop contributions to $V_3$ are very important
and, in the SM,  lead to a considerable strengthening of the PT
(fig. 1).\\
(iv) A delay in the PT towards lower $T_c$ strengthens the PT.
\\
(v) The rescaled sphaleron factor $A$ mentioned after Eq.\ (1) 
may be increased in
some variants.\\
(vi) Models with a tree level $\phi^3$-type term (like in NMSSMs)
may be interesting.

Here we  concentrate on the first point and argue
\cite{18}-\cite{23}
that in the Minimal Supersymmetric SM (MSSM) 
the $\phi^3$-type term in the effective potential is increased
if $\tilde{t}_R$, the superpartner of the right handed top,
is rather light. This enhancement is due to a diagram as shown below eq. (\ref{eq5})
but now with a stop in the loop. Its coupling to the $\phi^2$
background is given by the Yukawa coupling $h_t^2$.
To strengthen this term $h_t$ should be large and the thermal mass
\be\label{6}
m^2_{u_3}\sim m^2_u+cT^2\ee
of $\tilde t_R$ 
should be small in the $\phi$=0-phase. The physical $\tilde t_R$ mass is
$m^2_{\tilde t_R}=m^2_u+m^2_{top}$, where $m^2_u$ is
the SUSY-breaking scalar mass$^2$ and $m_{top}\sim h_t^2 \phi^2$. 
In renormalisation group equations 
starting with some $m^2_u$ at
the Planck scale one indeed
observes a decrease in $m^2_u$ much stronger than for other particles.
Large $h_t$ means rather small $\tan \beta=v_1/v_2$. It is also
convenient to make  one Higgs doublet heavy postulating
a large axial Higgs mass $m_a$. The Higgs mass $m_H$ is then fixed
by $\tan\beta$ and $m_a$.

If now the SUSY-partner particles and one Higgs doublet combination
are heavy enough, they can be ``integrated out'',
and one ends up at the same type of 3-dimensional
theory (3) but with different
relations between 4- and 3-dimensional parameters \cite{10, 21, 22,
23}. It turns out that one can get $v(T_c)/T_c> 1$
for $m_H\lsi 70/75$ GeV (fig. 5).
\begin{figure}
\centering
\epsfig{file=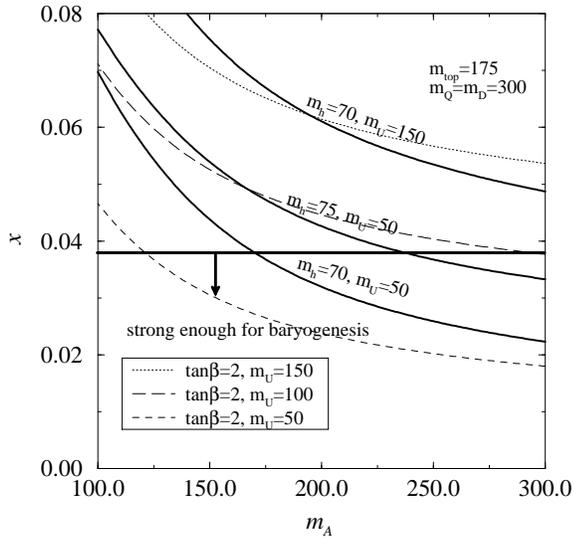, width=8cm}
\caption{(from ref. [28]). The parameter $x=\lambda_3/
g_3^2$ dependent on the axial mass $m_A$, the Higgs mass
$m_k$ (resp. $\tan\beta$) and the SUSY breaking $m_u$ in the MSSM.}
\end{figure}
One can even go further and discuss a  very small or even
negative $m^2_u$. However, to do this properly \cite{24}, the
stop field $U =\tilde{t}_R$ cannot be integrated out. It has to be included
in the 3-dimensional Lagrangian adding to (3) a term
\be\label{eq7}
L_3^{stop}=\frac{1}{4}{G^{A}_{ik}}^2+(D_i^sU)^+(D_i^sU)
+m^2_{u_3}U^+U+\lambda_{U_3}(U^+U)^2+\gamma_3H^+HU^+U\ee
where $G$ is the SU(3) Yang-Mills field strength and $D_i^s$ the
color covariant derivative, $m^2_{u_3}$ and $\lambda_{u_3}$ are the
mass$^2$ and the coupling like in (3), and the last term is a
mixing with, at tree level, $\gamma_3\sim Th^2_t\sin^2\beta$.
We have calculated the corresponding perturbative potential.
2-loop effects are important since \newpage only in this order gluon
and Higgs exchange in a $\tilde t_R$-loop come into play.
\begin{center}
\epsfig{file=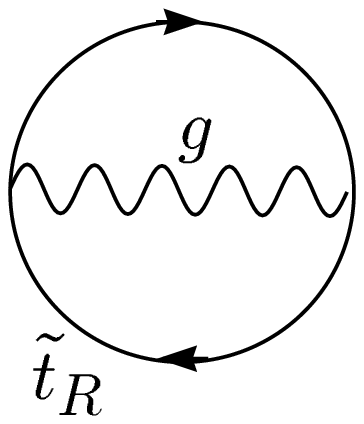,width=3cm}
\end{center}
This
leads to a much stronger PT. Going with $-m^2_{u_3}=\tilde m^2_u$ to
70 GeV, the upper limit for $m_H$ to obtain $v(T_c)/T_c>1$
becomes $m_H\leq 100$ GeV (fig. 6), still avoiding
an unstable Higgs vacuum
at $T=0$. Here we have put $\tilde A_t=0$ (see [24] for its definition)
for simplicity and in
order to obtain maximal effect.
\begin{figure}[ht]
\epsfig{file=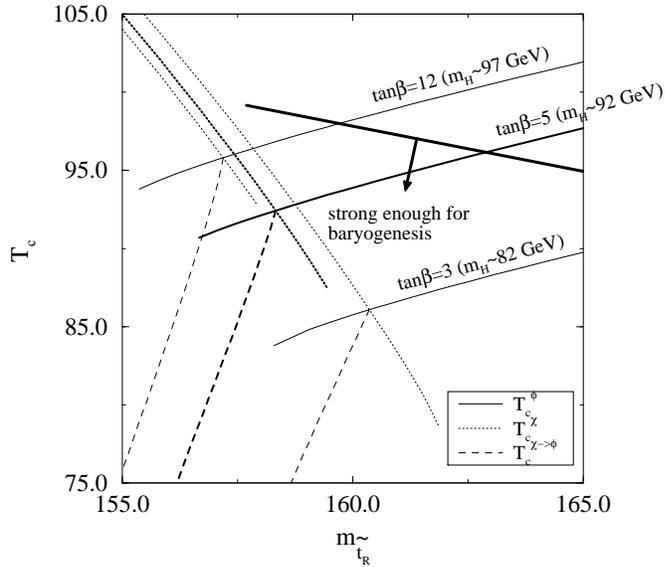,width=9cm}
\caption{(from ref. [31]) The critical temperatures
$T_c$ of the three transitions $0\to\phi,\ 0\to \chi,\ \chi\to
\phi$ for $\tan\beta=3,$ 12 (thin lines) and $\tan\beta$=5 (thick lines).
The two-stage transition would occur to the left of the crossing
point of the three critical curves. Also the boundary
$x\lsi 0.04$ for strongly first-order
PT $0\to\phi$ is indicated.}
\end{figure}
\begin{figure}
\vspace*{-1cm}
\begin{minipage}{\textwidth}
\epsfig{file=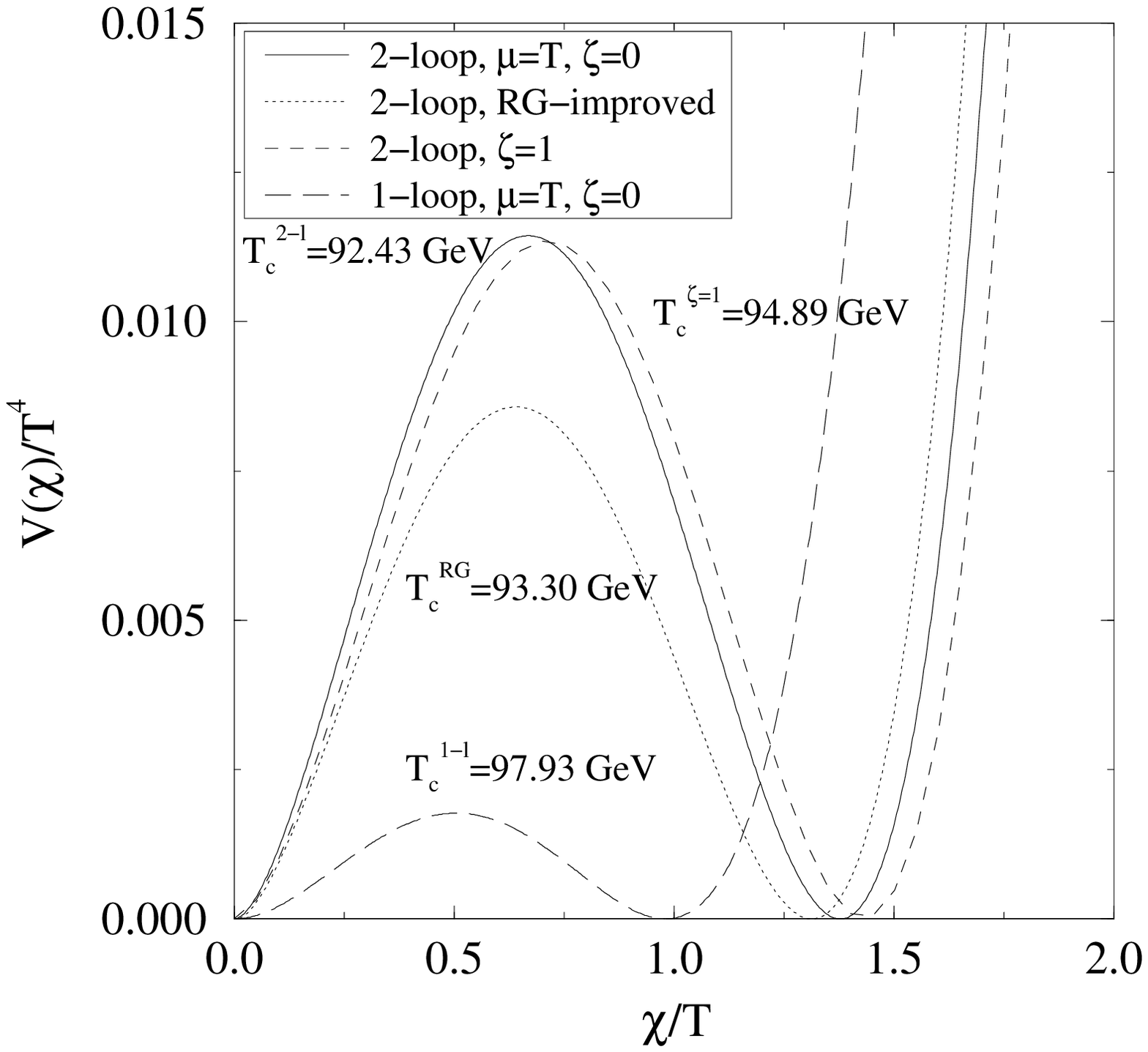,width=9cm}\vspace*{-1cm}
%
\epsfig{file=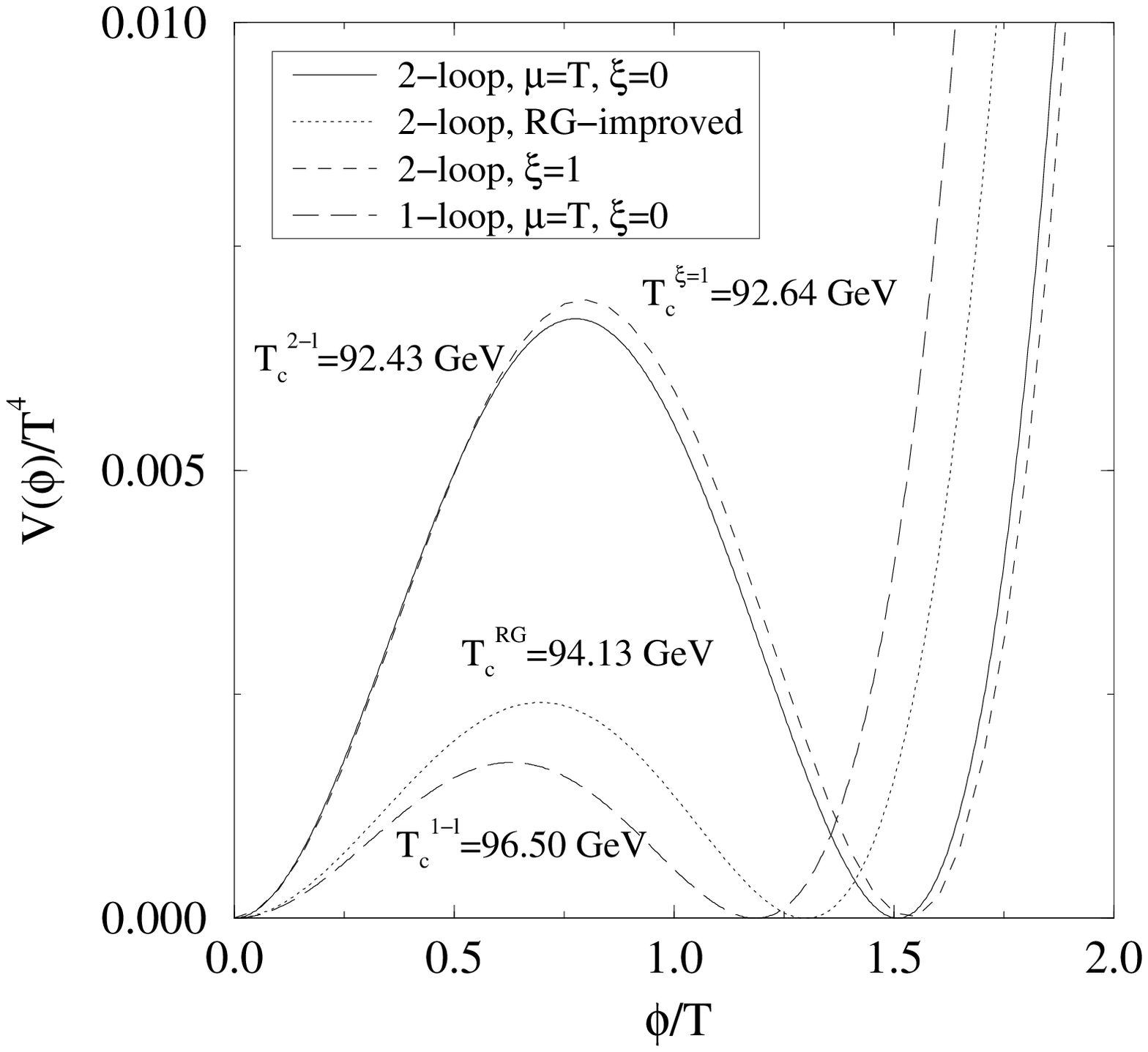,width=9cm}
\caption{(from ref. [31]) The 1- and 2-loop effective potentials
in $<$$U_3$$>$=$\chi$ and $<$$H$$>$=$\phi$ directions $(\tan\beta=5)$. Here
$m_{\tilde t_R}=158.3$ GeV is chosen such that the cricital
temperatures $T_c^{2-loop}$ of the two-phase transitions are
equal.}
\end{minipage}
\end{figure}

Most interestingly we also can obtain \cite{24} a two-stage
(figs. 6, 7)
PT in a certain range of $m_{\tilde t_R}\sim 155-160$ GeV. There is
a first-order PT to the ``colored'' minimum (strictly speaking,
in a gauge theory $<U^+U>\not=0$ and not $<U>\neq 0$!)
and then at lower $T$ a transition to the Higgs vacuum.
This delays the second $B$-asymmetry generating PT towards
lower $T$ and thus increases \cite{24} $v(T)/T$. But now one has
to make sure that the transition rate does not become too small.
Fortunately, in the
intermediate phase the sphaleron is Boltzmann-unsuppressed contrary
to the 2-Higgs-two-stage PT considered previously \cite{25}
and thus allows strong $B$-violation.

All this is based on perturbation theory and should be checked
by lattice calculations. Experience tells us that
perturbative calculations are o.k.
for a strongly first-order PT. Observing gauge
and $\mu$-dependence in our calculation \cite{25} (fig. 7)
as well as noting the
steep tree level potential between the minima, doubts are,
however, allowed.

In conclusion of the last part it can be said that the MSSM
variant of the SM allows a strongly first-order PT for Higgs
masses as large as 95 GeV. Even a two-stage PT is possible.
Given such a strong PT it is of interest to develop further
the machinery of producing $B$-asymmetry in front of expanding
bubbles which also requires the discussion of $CP$ violation in the
model.

I would like to thank D. B\"odeker, H. G. Dosch, P. John,
J. Kripfganz,  M. Laine and A. Laser
for enjoyable collaborations on various topics
reviewed in the report and also B. Bergerhoff, O. Philipsen
and C. Wetterich for useful discussions.

\end{document}